\documentclass[aps,prb,twocolumn, draft, showpacs,intlimits,amsmath,amssymb,floats]{revtex4-1}
\usepackage{bm}
\usepackage[final]{graphicx}
\usepackage{epsfig}

\begin{document}

\title{Doping Evolution of Oxygen $K$-edge X-ray Absorption Spectra in Cuprate Superconductors}

\author{C.-C. Chen$^{1,2}$}
\author{M. Sentef$^{1}$}
\author{Y. F. Kung$^{1,3}$}
\author{C. J. Jia$^{1,4}$}
\author{R. Thomale$^{3,5,6}$}
\author{B. Moritz$^{1,7,8}$}
\author{A. P. Kampf$^9$}
\author{T. P. Devereaux$^{1}$}

\address{$^1$Stanford Institute for Materials and Energy Sciences, SLAC National Accelerator Laboratory, Menlo Park, California 94025, USA}
\address{$^2$Advanced Photon Source, Argonne National Laboratory, Lemont, Illinois 60439, USA}
\address{$^3$Department of Physics, Stanford University, Stanford, California 94305, USA}
\address{$^4$Department of Applied Physics, Stanford University, Stanford, California 94305, USA}
\affiliation{$^5$Institut de th\'eorie des ph\'enom\`enes physiques, \'Ecole Polytechnique F\'ed\'erale de Lausanne (EPFL), CH-1015 Lausanne}
\affiliation{$^6$Institute for Theoretical Physics and Astrophysics, University of W\"urzburg, D 97074 W\"urzburg, Germany}
\address{$^7$Department of Physics and Astrophysics, University of North Dakota, Grand Forks, North Dakota 58202, USA}
\address{$^8$Department of Physics, Northern Illinois University, DeKalb, Illinois 60115, USA}
\address{$^9$Center for Electronic Correlations and Magnetism, Theoretical Physics III, Institute of Physics, University of Augsburg, 86135 Augsburg, Germany}

\date{\today}

\begin{abstract}
We study oxygen $K$-edge x-ray absorption spectroscopy (XAS) and investigate the validity of the Zhang-Rice singlet (ZRS) picture in overdoped cuprate superconductors. Using large-scale exact diagonalization of the three-orbital Hubbard model, we observe the effect of strong correlations manifesting in a dynamical spectral weight transfer from the upper Hubbard band to the ZRS band. The quantitative agreement between theory and experiment highlights an additional spectral weight reshuffling due to core-hole interaction. Our results confirm the important correlated nature of the cuprates and elucidate the changing orbital character of the low-energy quasi-particles, but also demonstrate the continued relevance of the ZRS even in the overdoped region.
\end{abstract}
\pacs{78.70.Dm, 74.25.Jb, 74.72.-h, 78.20.Bh}
\maketitle

\section{Introduction}
\begin{figure}[t!]
\includegraphics[width=\columnwidth]{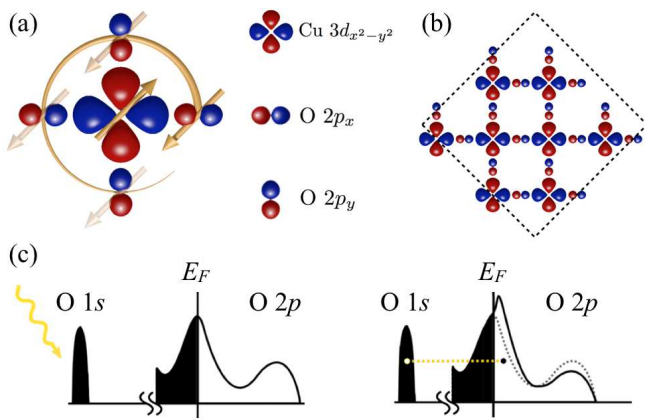}
\caption{
(Color online) (a) Cartoon picture of a Zhang-Rice singlet formed by a copper $3d_{x^2-y^2}$ hole hybridized with a hole on its neighboring oxygen $2p_{x,y}$ orbitals.
(b) The Cu$_8$O$_{16}$ cluster with periodic boundary conditions considered in the exact diagonalization calculations.
(c) Schematic of the oxygen $K$-edge x-ray absorption process, highlighting additional spectral weight reshuffling due to core-hole interaction.
}\label{fig:ClusterandZRS}
\end{figure}

Unraveling the nature of low-energy quasi-particles in cuprate superconductors is crucial to understanding their unconventional superconducting mechanism. Despite two decades of studies, however, the minimal model for describing the quasi-particle band which emerges upon doping remains unclear. Previous theoretical work has focused on the single-band Hubbard model\cite{singleband_Hubbard} or the $t-J$ model obtained by projecting out doubly-occupied charge configurations.\cite{tJ_Spalek, ZRS_tJ} In these down-folded Hamiltonians, the fundamental quasi-particle has been assigned to the so-called Zhang-Rice singlet (ZRS):\cite{ZRS_tJ, ZRS_George} a locally bound $d^9$ copper $3d_{x^2-y^2}$ hole hybridized with a doped ligand hole ($L$) distributed on the planar oxygen $2p_{x,y}$ orbitals [Fig. \ref{fig:ClusterandZRS}(a)].

The relevance of the ZRS in cuprate materials is supported by various spectroscopies: spin-resolved photoemission confirms the singlet character of the first ionization state,\cite{ZRS_SpinPhotoemission} and other probes show that doped holes in undoped or lightly-doped cuprates reside primarily on oxygens.\cite{PES_ZXShen, XAS_Tranquada, EELS_Nucker, Compton} The doping evolution of oxygen content has been studied extensively by core-level x-ray absorption spectroscopy (XAS).\cite{XAS_CTChen,XAS_CTChen_pz, XAS_Pellegrin}
In the insulating parent compounds, early oxygen $K$-edge XAS experiments (oxygen core-electron $1s\rightarrow 2p$ transition) show that below the main absorption edge a single ``pre-peak" exists near 530 eV,\cite{XAS_CTChen} attributed to excitations into the upper Hubbard band (UHB). This nonzero projection of the UHB onto the oxygen XAS spectra mainly results from a strong hybridization between copper and oxygen, and the mixing of $d^9$ and $d^{10}L$ character in the ground state.
Upon hole-doping, a lower-energy ($d^9L$) peak emerges near 528.5 eV and grows linearly with hole concentration, while the intensity of the higher-energy UHB decreases.
Polarized x-ray studies indicate that the lower-energy peak in the underdoped regime shows almost no oxygen $2p_z$ content, but is dominated by planar oxygen $2p_{x,y}$ contributions, \cite{XAS_CTChen_pz, XAS_Pellegrin} lending further credence to the ZRS picture.

However, recent measurements report that while the low-energy ZRS weight increases linearly with doping over a wide doping range, it deviates from a linear trend and exhibits a weak doping dependence beyond $\sim20\%$.\cite{XAS_saturation_Schneider, XAS_Peets} It was suggested that this ``anomalous" behavior implies the inapplicability of the single-band Hubbard model and a breakdown of the ZRS picture.\cite{XAS_Peets}
But cluster dynamical mean-field theory (DMFT) calculations argue that the single-band Hubbard model could explain the experimental finding, by showing a change in slope with doping of the integrated unoccupied density of states (DOS) beyond a certain doping level.\cite{DMFT_Comments}
On the other hand, calculations for the more sophisticated three-orbital (copper-oxygen) Hubbard model with single-site DMFT find that the low-energy weight in the oxygen-projected DOS simply increases linearly with doping; the result contrasts with experiment and challenges even the validity of the three-orbital model.\cite{XAS_Wang}
Currently, the debate continues and questions remain.\cite{XAS_Peets, XAS_Wang, DMFT_Comments, DMFT_Liebsch, XAS_Reply_1, XAS_Reply_2} Can oxygen $K$-edge XAS be fully described by the Hubbard models? Are these model Hamiltonians and the ZRS picture suitable for the cuprates in the overdoped region?

In this work, we investigate these questions by tracking the doping evolution of photoemission and oxygen $K$-edge XAS spectra in the three-orbital Hubbard model. Using large-scale exact diagonalization, we treat many-body correlations exactly and include the effect of core-hole interaction explicitly, which is essential in the proper evaluation of the XAS cross sections. Our calculations reproduce the experimental data in a quantitative way and explain the crossover or apparent saturation in ZRS intensity in the overdoped regime, highlighting the dynamical spectral weight transfer from the UHB to the ZRS band expected for strongly correlated systems. The slope of the ZRS weight versus doping further decreases due to the presence of core-hole interaction. These results confirm the correlated, multi-orbital nature of the low-energy quasi-particles in the cuprates, but also indicate that the ZRS picture may indeed survive to high doping levels, albeit with an evolving orbital character.

\section{Methods}
We consider the three-orbital Hubbard model on a square lattice of planar CuO$_2$ plaquettes containing copper $3d_{x^2-y^2}$ and oxygen $2p_{x,y}$ orbitals:\cite{Mattheiss, Varma_threeband, Emery_model} 
\begin{eqnarray}
H&=&-\sum_{\langle i,j\rangle\sigma} t^{ij}_{pd}(d^\dagger_{i\sigma} p_{j\sigma}+h.c.)
-\sum_{\langle j,j'\rangle\sigma} t^{jj'}_{pp}(p^\dagger_{j\sigma} p_{j'\sigma}+h.c.)
\nonumber\\
&+&\sum_{i\sigma} \epsilon_d  n^d_{i\sigma}  + \sum_{j\sigma} \epsilon_p  n^p_{j\sigma}+ \sum_i U_d n^d_{i\uparrow} n^d_{i\downarrow}+ \sum_j U_p n^p_{j\uparrow} n^p_{j\downarrow}.
\nonumber
\end{eqnarray}
Here $d^\dagger_{i\sigma}$ ($p^\dagger_{j\sigma}$) creates a hole with spin $\sigma$ at a copper site $i$ (oxygen site $j$), and $n^d_{i\sigma}$ ($n^p_{j\sigma}$) is the copper (oxygen) hole number operator.
The first two terms of the Hamiltonian represent the nearest-neighbor copper-oxygen and oxygen-oxygen hybridization. The hopping integrals ($ t^{ij}_{pd}$ and $t^{jj'}_{pp}$) can change sign depending on the phases of the overlapping wavefunctions [Fig. \ref{fig:ClusterandZRS}(a)].\cite{Emery_model, threeband_VCA} The third and fourth terms are the copper and oxygen site energies, and the last two terms are the on-site Hubbard interactions on copper and oxygen, respectively. We use the following parameters (in units of eV): $\epsilon_p-\epsilon_d=3.23$; $U_d=8.5$, $U_p=4.1$; $|t_{pd}|=1.13$, $|t_{pp}|=0.49$. These parameters have been employed to study the cuprate material La$_2$CuO$_4$ and produce various spectral features in good agreement with experiment.\cite{Ohta_parameters, LCO_RIXS}

Using exact diagonalization for a Cu$_8$O$_{16}$ cluster with periodic boundary conditions [Fig. \ref{fig:ClusterandZRS}(b)], we obtain the ground state and calculate the photoemission and XAS cross sections for four doping levels: 0\% (undoped), 12.5\% (underdoped), 25\% (overdoped), and 37.5\% (heavily overdoped).
To calculate the spectral functions, we exploit translational symmetries of the problem and diagonalize the Hamiltonian matrices in momentum space. For oxygen $K$-edge XAS, the presence of local core-hole interaction breaks translational symmetry, and we perform the calculations in real space with a larger matrix size.
The largest Hilbert space considered in this work contains $\sim5.7\times 10^9$ basis states.\cite{EDalgorithm}

\section{Results and Discussion}

\subsection{Spectral Functions and Photoemission Spectra}

\begin{figure}[t!]
\includegraphics[width=\columnwidth]{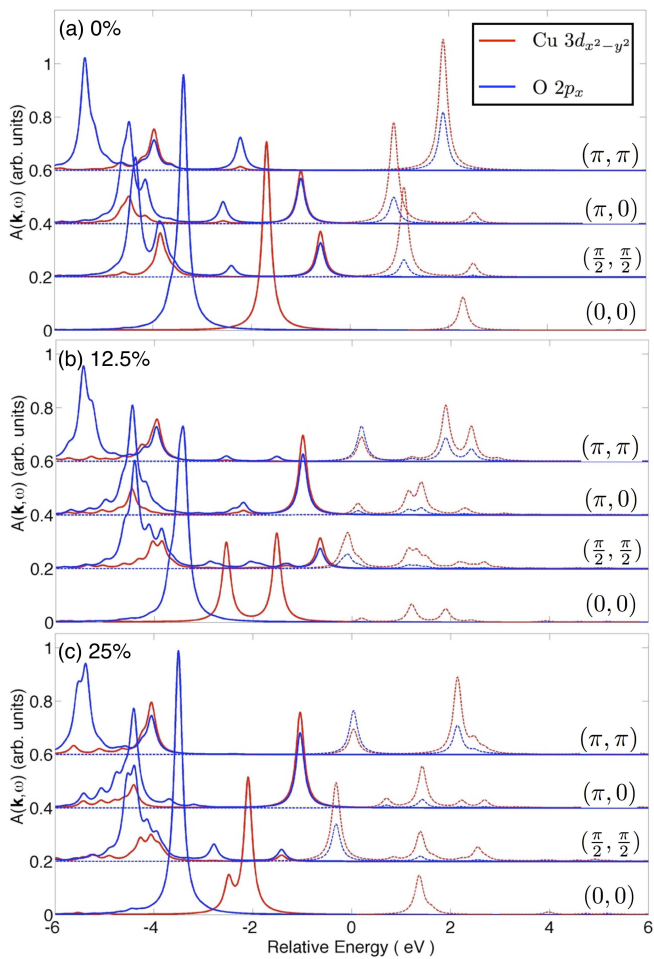}
\caption{
(Color online) Spectral functions calculated using exact diagonalization on a Cu$_8$O$_{16}$ cluster at (a) half-filling, (b) 12.5$\%$, and (c) 25$\%$ hole dopings. The ground state energy of the valence electrons at the corresponding filling is defined as zero energy. The solid and the dashed lines represent the electron occupied and unoccupied states, respectively. The spectra are broadened with a 0.1 eV Lorentizian.
}\label{fig:AKW}
\end{figure}

Oxygen $K$-edge XAS promotes $1s$ electrons into empty $2p$ states [Fig. \ref{fig:ClusterandZRS}(c)], and the resulting spectrum is related to the unoccupied oxygen projected-DOS measured in inverse photoemission spectroscopy (IPES). To capture the doping evolution of electronic structures, we first discuss the orbitally-resolved spectral functions in Fig. \ref{fig:AKW}, where energy zero is defined by the valence-electron ground-state energy at the corresponding filling. At $0\%$ doping [Fig. \ref{fig:AKW}(a)], the spectra show an indirect gap of size $\sim$1.5 eV, related to charge-transfer excitation of moving a hole from copper to oxygen.\cite{CTinsulator_ZSA} Without doping, the first electron removal state appears at $(\pi/2,\pi/2)$, while the first electron addition state occurs at $(\pi,0)$. This indirect gap and its size are consistent with angle-resolved photoemission (ARPES) experiments.\cite{indirectgap_NCCO} In an indirect gap system, optical conductivity measurements would obtain a gap larger than the actual band gap, as no net momentum is transferred in the process. We find an optical gap $\sim1.7$ eV, also in agreement with the observed gaps (1.5-2.0 eV) in insulating cuprate parent compounds.\cite{OPC_Tokura, OPC_Cooper, OPC_Uchida, OPC_Falck}

Upon hole-doping, low-energy peaks above the Fermi level emerge especially at momentum points along the magnetic Brillouin zone boundary.
This emergent band [near zero energy in Figs. \ref{fig:AKW}(b) and  \ref{fig:AKW}(c)] consisting of both copper and oxygen are associated with the ZRS; its orbital composition, spectral intensity, and energy position show strong momentum-dependence. In particular, the ZRS band contains vanishing oxygen weight at the $\Gamma$ point, due to a cancellation in the phase factor of the ZRS wavefunction,\cite{ZRSweight_Unger, ZRSweight_Pothuizen, ZRSweight_Yin} which is well-defined only away from the $\Gamma$ point.
This is not captured in the single-band model and stresses the need for considering the multi-orbital nature in describing the ZRS doping evolution at a quantitative level.

On the occupied side [solid lines in  Fig. \ref{fig:AKW}], the spectral features at $(\pi,0)$ and $(\pi/2,\pi/2)$ show distinct doping dependences: While the weight of the highest occupied state at $(\pi/2, \pi/2)$ gradually decreases upon hole-doping, the intensity increases for the highest occupied state at $(\pi,0)$. Moreover, the first electron removal state changes from momentum $(\pi/2,\pi/2)$ in the underdoped region to $(\pi,0)$ on the overdoped side,\cite{ARPES_SCOC_Kim, ARPES_LCO_Ino} reminiscent of the observations from ARPES.\cite{Waterfall_footnote} This doping dependence cannot be accounted for by a simple chemical-potential shift in a rigid-band model, but is a manifestation of spectral weight redistribution due to correlation effects.\cite{dynamical_weight_singleband, dynamical_weight_Eskes, XAS_singleband, dynamical_weight_Meinders, dynamical_weight_PES}

The occupied bands between roughly $-3$ to $-5$ eV are related to oxygen bands dispersing with $t_{pp}$\cite{NB_Mizuno} and the Zhang-Rice triplet.\cite{ZRT_Learmonth} In real cuprate materials, other orbitals (such as copper $3d_{3z^2-r^2}$ and non-bonding oxygens) also contribute to the spectral weight in this energy range.\cite{DOS_DMFT_Weber} Below -6 eV [not shown], we find a highly incoherent band containing mostly copper character, associated with the lower Hubbard band. These bands at deeper binding-energies show a relatively weak doping dependence in the doping range considered in this study.

 \begin{figure}[t!]
\includegraphics[width=\columnwidth]{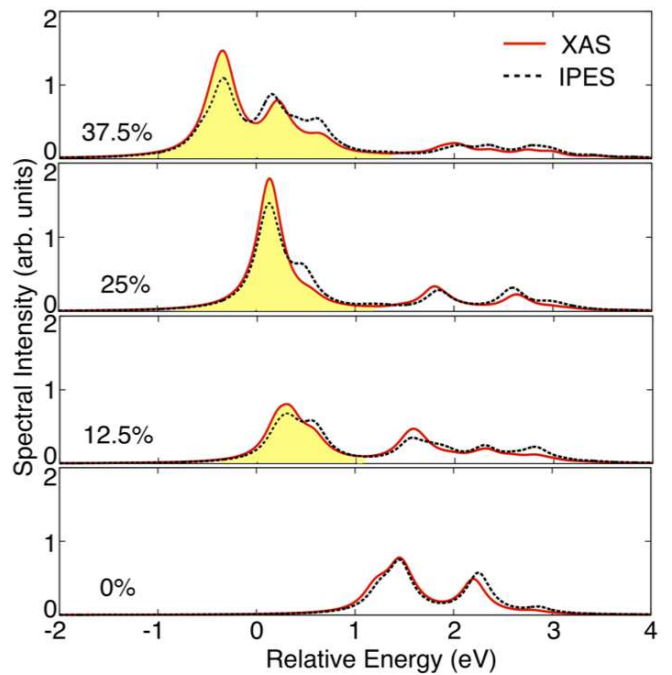}
\caption{
(Color online) Calculations of oxygen $K$-edge XAS (solid red lines) and IPES measurements of unoccupied oxygen-projected DOS (black dotted lines) at various dopings.
The yellow shaded area represents the emergent low-energy ZRS band.
The XAS and IPES spectra are lined up in energy by their peak maximum positions.
}\label{fig:OKXAS}
\end{figure}

\subsection{Oxygen $K$-edge X-ray Absorption Spectra}

\begin{figure}[t!]
\includegraphics[width=\columnwidth]{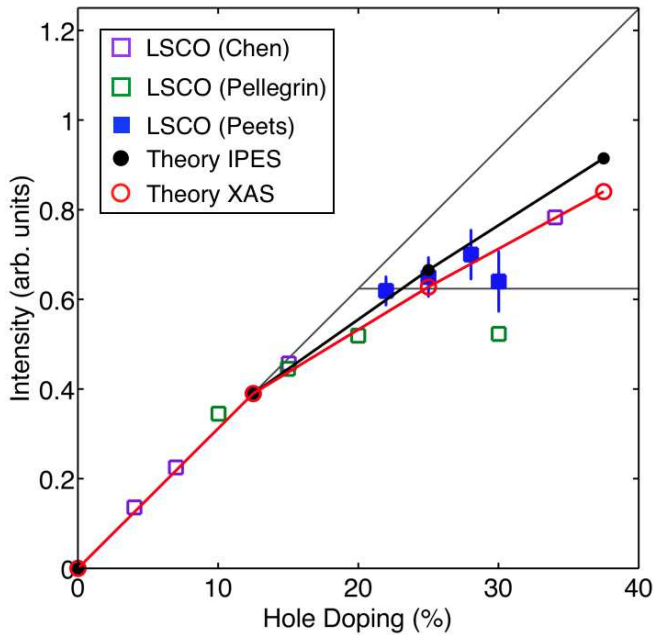}
\caption{
(Color online) Doping dependence of the ZRS spectral weight. The La$_{2-x}$Sr$_x$CuO$_4$ (LSCO) XAS data are from Refs. \onlinecite{XAS_CTChen} (Chen), \onlinecite{XAS_Pellegrin} (Pellegrin), and \onlinecite{XAS_Peets} (Peets). The XAS theory (red circle) is obtained by integrating the yellow shaded area in Fig. \ref{fig:OKXAS}. The black circle represents the ZRS weight from the IPES calculation of the unoccupied DOS.
Both theory curves are rescaled by normalizing their respective 12.5\% ZRS weight to the straight line suggested by experiment, \cite{XAS_Peets} and the ZRS intensity at $0\%$ doping is defined as zero .
}\label{fig:XASZRSweight}
\end{figure}

We next discuss in Fig.  \ref{fig:OKXAS} the oxygen $K$-edge XAS calculations, which include an additional oxygen $1s-2p$ core-hole interaction ($\sum_{i\sigma\sigma'} U_c n^{1s}_{i\sigma} n^{2p}_{i\sigma'}$) with $U_c=6$ eV.\cite{coreholestrength, UQonXAS_Eskes, OKXAS_1DUQ_Okada}
The spectral peaks at $0\%$ doping are closely related to the UHB [near 1-2 eV in Fig. \ref{fig:AKW}], which upon doping becomes more incoherent and shifts its weight to the lower-energy ZRS band.
The energy separation between these two bands increases systematically with doping: the UHB disperses toward higher energy and the ZRS band moves in the opposite direction, in agreement with experiment.\cite{XAS_CTChen}
Figure  \ref{fig:OKXAS} also shows the IPES calculations of the oxygen unoccupied DOS.
The $K$-edge XAS lineshape resembles closely that of IPES, because the only core-hole interaction is a monopole term forming a simple charge density attraction  due to the isotropic $1s$ orbital.\cite{LedgeXAS}
However, the two spectra still vary quantitatively [Fig. \ref{fig:XASZRSweight}].\cite{RIXS_Jia}

In Fig. \ref{fig:XASZRSweight} we compare our theory with oxygen $K$-edge XAS experiments on La$_{2-x}$Sr$_x$CuO$_4$ (LSCO).\cite{XAS_Peets} LSCO has a relatively simple crystal structure and is suitable for systematic studies of electronic properties over a wide doping range. As shown in Fig. \ref{fig:XASZRSweight}, the experimental ZRS weight increases roughly linearly with doping in the underdoped region.
Near optimal doping, the rate of increase changes and exhibits a weaker doping dependence (smaller slope). It was suggested that the ZRS intensity may even saturate on the overdoped side ($\geq20\%$).\cite{XAS_saturation_Schneider, XAS_Peets}

For comparison, in Fig. \ref{fig:XASZRSweight} we also plot the ZRS weight of the XAS and IPES calculations.
We define the ZRS intensity as zero at $0\%$ doping and rescale both theory curves by normalizing their respective 12.5\% ZRS weight to the straight line suggested by experiment.\cite{XAS_Peets}
Neither calculation shows a saturation, but only a linear increase at low dopings and a change of slope at a higher doping level.
This crossover in slope is related to the spectral weight transfer from the UHB to the ZRS band, which can be illustrated by considering the single-band Hubbard model in the atomic limit.\cite{dynamical_weight_singleband, dynamical_weight_Eskes, XAS_singleband, dynamical_weight_Meinders, dynamical_weight_PES} On a half-filled $N$-site lattice, the electron removal spectrum (associated with singly occupied states) and addition spectrum (associated with the unoccupied UHB) both have a spectral weight equal to $N$. Upon doping $x$ holes, there are $N-x$ singly occupied states, and the UHB weight also becomes $N-x$. Due to the conserved sum of the occupied and unoccupied states ($2N$), the low-energy spectral weight which emerges upon doping $x$ holes is thereby increased by 2$x$. In contrast, for an uncorrelated system the emergent spectral weight equals $x$ when doped with $x$ holes.
We also note that an effective weakening in correlation strength and spectral weight transfer upon doping is absent in the $t-J$ model, which always behaves as a single-band Hubbard model in the atomic limit.\cite{dynamical_weight_Eskes, OKXAS_Bansil}


Although we do not find an apparent saturation of the ZRS weight with doping, our XAS calculation seems to capture well the experimental findings. In particular, the XAS theory curve matches the experimental data within error bars near optimal doping and in the overdoped regime; our theory also predicts the doping dependence of the heavily overdoped ($x\sim0.35$) LSCO sample.
Compared with IPES ($U_c=0$), the core-hole attraction ($U_c=6$ eV) effectively overcomes the energy cost of double occupancy and shifts roughly an additional $\sim10\%$ of the UHB weight to lower energies.
This effect is more prominent at low doping where a substantial UHB weight still remains, leading to a more pronounced change of slope in ZRS intensity in XAS.
The IPES  calculations \emph{underestimate the low-energy spectral weight} at low dopings and \emph{overestimate the slope} of the ZRS intensity with doping in the overdoped region.
We note that Liebsch\cite{DMFT_Liebsch} and also Peets \emph{et al.}\cite{XAS_Reply_2} have pointed out the importance of choosing a proper energy integration window for the ZRS band when comparing theory and experiment. Here we use a $\sim$2 eV energy window whose upper limit changes with doping and resides in the minimum between the centroids of the UHB and the ZRS band (see Fig. \ref{fig:OKXAS}), consistent with experiment.\cite{XAS_Peets, XAS_Reply_2} 

As mentioned previously, the doping evolution of the low-energy ZRS spectral weight has also been studied by various theoretical techniques.\cite{XAS_Wang, DMFT_Comments, DMFT_Liebsch, XAS_singleband, OKXAS_Bansil, OKXAS_Okada, XAS_FEFF} In particular, single-site DMFT using the three-orbital model shows a linear increase with doping of the low-energy weight.\cite{XAS_Wang}
Cluster DMFT of the single-band model demonstrates a change in slope of the ZRS weight beyond a certain doping level,\cite{DMFT_Liebsch, DMFT_Comments}
which resembles the behavior of our calculations for the unoccupied DOS in the three-orbital model.
However, the slope on the overdoped side will be overestimated in the single-band case,\cite{dynamical_weight_Meinders, XAS_singleband, DMFT_Comments} as there the UHB completely disappears close to 100\% hole doping, while in the three-orbital model the $d^8$ double occupancy stays nonzero. 
Therefore, if the multi-orbital nature of the problem is neglected, an effective doping-dependent Coulomb repulsion needs to be imposed in the single-band model in order to quantitatively capture the spectral weight transfer in the overdoped region.\cite{OKXAS_Bansil}

\section{Conclusion}

To summarize, our exact diagonalization calculation of the three-orbital Hubbard model agrees with the XAS experiments at a quantitative level, which relies crucially on the combined effect of correlation-induced spectral weight transfer and an additional spectral weight redistribution due to core-hole interaction.
The change of slope in ZRS intensity as a function of doping manifests clearly the presence of strong electronic correlation.
We also find a strong doping and momentum dependence of the orbital composition in the ZRS band, which cannot be fully described by the single-band Hubbard model especially in the overdoped region.
Our results potentially confirm the continuation of the ZRS picture even in overdoped cuprates. Further measurements would be helpful to illuminate the nature of the emergent low-energy quasi-particles in cuprate superconductors.

\acknowledgments
The authors acknowledge discussions with M. A. van Veenendaal, J.  Fernandez-Rodriguez, and C.-Y. Mou.
This work is supported by the U.S. Department of Energy (DOE), Basic Energy Sciences, Office of Science, under Contracts No. DE-AC02-76SF00515 and No. DE-AC02-06CH11357.
C.C.C. is supported by the Aneesur Rahman Postdoctoral Fellowship at ANL.
Y.F.K. was supported by the Department of Defense (DoD) through the National Defense Science {\&} Engineering Graduate Fellowship (NDSEG) Program.
C.J.J. is supported by the Stanford Graduate Fellowships. 
R.T. is supported by an SITP Fellowship at Stanford University and  SPP 1458/1.
A.P.K. acknowledges support from the DFG through TRR 80.
The simulations were performed on the Hopper peta-flop Cray XE6 system at NERSC, supported by the U.S. DOE under Contract No. DE-AC02-05CH11231.


\begin{thebibliography}{arprev}
\bibitem{EDalgorithm}
To obtain the ground state, we use both the Lanczos tridiagonalization technique and the Parallel ARnold PACKage based on implicitly restarted methods.\cite{PARPACK} We adopt a block-checkerboard matrix decomposition with a parallel computing scheme implemented on a distributed memory system. By partitioning both the rows and columns of a matrix, each computing thread no longer stores a full vector, and the matrix diagonalization is not limited by the total memory per computing node.
\bibitem{Waterfall_footnote}
We note that although our spectral functions exhibit  several non-trivial momentum and doping dependences in agreement with ARPES measurements, due to the limitation of the calculations, we are unable to track the band dispersion in fine details at momentum transfers unavailable in our finite-size cluster, such as the $(\pi/4,\pi/4)$ point near where the so-called ``waterfall anomaly" occurs.\cite{Waterfall_expt1, Waterfall_expt2, Waterfall_expt3, Waterfall_expt4, Waterfall_expt5, Waterfall_theory1, Waterfall_theory2, Waterfall_theory3, Waterfall_theory4, Waterfall_theory5}
\bibitem{coreholestrength}
In copper $K$-edge XAS, the $1s$ core-hole interaction is found to be slightly larger than the on-site Hubbard interaction based on the resonant energies of the well- and poorly-screened states. For oxygen $K$-edge XAS, the oxygen Hubbard repulsion $U_p$ is estimated to be 4-5 eV, and a core-hole strength $U_c\sim$5-6 eV is commonly used.
\end{thebibliography}
\end{document}